# Silicon nitride stress-optic microresonator modulator for optical control applications

**JIAWEI WANG,[1,*] KAIKAI LIU,[1] MARK W. HARRINGTON,[1] RYAN Q. RUDY,[2] AND DANIEL J. BLUMENTHAL[1]**

[1]*Department of Electrical and Computer Engineering, University of California, Santa Barbara, Santa Barbara, CA 93106, USA*
[2]*U.S. Army Research Laboratory, Adelphi, Maryland 20783, USA*
*\*jiawei_wang@ucsb.edu*

**Abstract:** Modulation-based control and locking of lasers, filters and other photonic components is a ubiquitous function across many applications that span the visible to infrared (IR), including atomic, molecular and optical (AMO), quantum sciences, fiber communications, metrology, and microwave photonics. Today, modulators used to realize these control functions consist of high-power bulk-optic components for tuning, sideband modulation, and phase and frequency shifting, while providing low optical insertion loss and operation from DC to 10s and 100s of MHz. In order to reduce the size, weight and cost of these applications and improve their scalability and reliability, modulation control functions need to be implemented in a low loss, wafer-scale CMOS-compatible photonic integration platform. The silicon nitride integration platform has been successful at realizing extremely low waveguide losses across the visible to infrared and components including high performance lasers, filters, resonators, stabilization cavities, and optical frequency combs. Yet, progress towards implementing low loss, low power modulators in the silicon nitride platform, while maintaining the planar, wafer-scale process compatibility has been limited. Here we report a significant advance in integration of a piezo-electric (PZT) actuated micro-ring modulation in a fully-planar, wafer-scale silicon nitride platform, that maintains low optical loss (0.03 dB/cm in a 625 μm resonator) at 1550 nm, with an order of magnitude increase in bandwidth (DC to 20 MHz 3-dB) and order of magnitude lower power consumption of 20 nW improvement over prior PZT modulators. The modulator provides a >14 dB extinction ratio (ER) and 7.1 million quality-factor (Q) over the entire 4 GHz tuning range, a tuning efficiency of 200 MHz/V, and delivers the linearity required for control applications with 65.1 dB·Hz$^{2/3}$ and 73.8 dB·Hz$^{2/3}$ third-order intermodulation distortion (IMD3) spurious free dynamic range (SFDR) at 1 MHz and 10 MHz respectively. We demonstrate two control applications, laser stabilization in a Pound-Drever Hall (PDH) lock loop, reducing laser frequency noise by 40 dB, and as a laser carrier tracking filter. This PZT modulator design can be extended to the visible in the ultra-low loss silicon nitride platform with minor waveguide design changes. This integration of PZT modulation in the ultra-low loss silicon nitride waveguide platform enables modulator control functions in a wide range of visible to IR applications such as atomic and molecular transition locking for cooling, trapping and probing, controllable optical frequency combs, low-power external cavity tunable lasers, quantum computers, sensors and communications, atomic clocks, and tunable ultra-low linewidth lasers and ultra-low phase noise microwave synthesizers.

## 1. Introduction

The silicon nitride ($Si_3N_4$) photonic integrated platform [1,2] delivers ultra-low waveguide losses and broad passive functionality across the visible through the IR wavelength range [3,4] in a planar wafer-scale, CMOS-compatible process. This platform has the potential to provide the benefits of integration, namely low-cost, portability, low-power, scalability, and enhanced reliability to a wide range of applications including quantum information sciences and applications [5,6], optical atomic clocks [7,8], precision metrology [9,10], atomic, molecular and optical [11], microwave photonics [12,13], fiber optic precision frequency distribution [14–16], and energy efficient communications [17]. A ubiquitous function among these applications is optical actuation and modulation, to perform tuning and control functions including wavelength shifting, sideband and sweep modulation, and phase shifting, with low optical insertion loss and modulation bandwidths from DC to 10s and 100s of MHz. Examples include laser locking to reference cavities and atomic transitions [18] and optical channel tracking in fiber communications [19]. Today, these systems use power-consuming bulk-optic electrooptic (EOM) [20] and acousto-optic (AOM) [21] modulators to implement control loops such as the PDH [22,23] and proportional, integration and derivative (PID) [18,24]. Devices and functions that will benefit from integrated, low power, active modulation include hybrid tunable lasers [25], optical frequency combs [26,27], Brillouin lasers [4,28,29], atomic and ion transition trapping, cooling and locking [7,30], laser stabilization [31], phase lock loops [14], and cryogenic applications [32,33]. Yet it has remained a challenge to transition these modulators to the silicon nitride platform while maintaining the ultra-low optical loss in a CMOS-compatible, wafer-scale process. Therefore, integrated control modulators that maintain the desirable properties of the silicon nitride platform, as well as low-power consumption, are needed.

There have been efforts to realize tuning and modulation in ultra-low loss silicon nitride photonics. Silicon nitride has a high third-order Kerr nonlinearity which is suitable for generating octave-spanning frequency comb [34], however, due to its centrosymmetric crystal structure, silicon nitride has a small second-order Pockels effect with a maximum electro-optic (EO) coefficient of 8.31±5.6 fm/V [35] and cannot make use of free-carrier modulation like silicon [36]. Conventional thermal tuning approaches support bandwidths up to 10 kHz [37] with 40 mW silicon nitride tuners [33]. Heterogeneously integration of nonlinear materials and silicon nitride waveguides introduce second-order nonlinearities for EO modulation, for example lithium niobate [38], ferroelectrics [39], and zinc oxide [40]. These methods offer modulation bandwidths above 1 GHz, however, they suffer from large optical losses, increased fabrication complexity, high power consumption and can have limited material optical wavelength range. The stress-optic effect offers a broad optical bandwidth, moderate electrical bandwidth, and low power consumption. Stress-optic actuated silicon nitride waveguides have been demonstrated using piezoelectric materials like aluminum nitride (AlN) [32,41] and lead zirconate titanate (PZT) [42,43]. AlN actuation utilizes acoustic resonant enhancement due to its relatively small piezoelectric coefficient (AlN: $e_{31,f} = 1.02 \ C/m^2$, PZT: $e_{31,f} = -18.3 \ C/m^2$ [44]) and tuning efficiency which is an order of magnitude weaker than that of PZT [45]. The under-etched PZT approach results in high loss, low Q, limited modulation bandwidth (< 1 MHz) and is difficult to make compatible with planar wafer-scale integration [43]. Progress has been made with planar processed PZT actuators including a Mach-Zehnder interferometer (MZI) phase modulator [42] with a modulation bandwidth of 629 kHz and high optical loss.

In this paper, we report demonstration of a low-power photonic integrated PZT actuated stress-optic microresonator modulator for optical control functions, realized in the CMOS-compatible ultra-low loss silicon nitride platform. The waveguide-offset, fully planar design achieves 0.03 dB/cm loss in a 625 μm radius resonator operating at 1550 nm, with a DC to 20 MHz modulation bandwidth, an order of magnitude improvement over prior PZT stress-optic modulators [43], and power consumption of 20 nW. The resonator has an intrinsic Q of 7.1 million, which is over 10 times larger than prior state of the art [43] and 14 dB ER over a 4 GHz tuning range. The modulator resonance tuning coefficient is measured to be 1.6 pm/V (200 MHz/V) and corresponding half-wave voltage-length products are $V_\pi L$ = 43 V·cm and $V_\pi L \alpha$ = 1.3 V·dB. The IMD3 SFDR is measured to be 65.1 dB·Hz$^{2/3}$ and 73.8 dB·Hz$^{2/3}$ at 1 MHz and 10 MHz, respectively. We demonstrate the use the of this modulator in two applications, first as a double sideband (DSB) modulator in a laser frequency stabilization PDH lock loop, and second as a laser frequency tracking filter. Since the modulation is based on stress-optic induced changes in the silicon nitride, this design will work with ultra-low loss visible to IR designs with only small waveguide design changes to operate in the visible [3]. This advance in ultra-low loss waveguide modulators will enable chip-level control to be integrated with other ultra-low loss silicon nitride components [1] for a wide range of visible to IR applications, including atomic and molecular quantum sensing, computing and communicaton, controllable optical frequency combs, low-power stabilized lasers, atomic clocks, and ultra-low phase noise microwave synthesizers.

An example in the silicon nitride platform that can employ this modulator is a frequency stabilized laser, illustrated in Fig. 1. With this potential application, the low-power tuning PZT actuator can be used to fine tune a silicon nitride dual-resonators vernier laser [25] and carrier lock the laser to an on-chip coil reference cavity [46] using the double sideband modulator and a PDH lock loop [22]. The stabilized laser output can be further modulated for locking to applications such as quantum and atomic applications, position and navigation, and coherent communications.

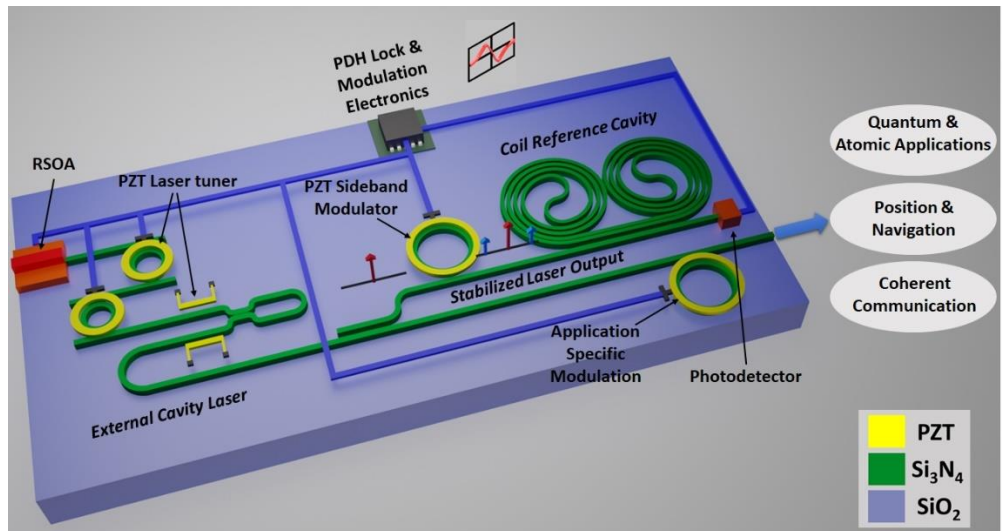

Fig. 1 Illustrative example application of PZT-actuated modulator for control functions in an external cavity laser (ECL). The PZT actuator can be used to tune and control the ECL, and generate sidebands to further stabilize the laser fluctuation by Pound-Drever Hall (PDH) locking to an ultra-high Q reference cavity. The stabilized output can be modulated (optional) based on specific applications.

## 2. Modulator Design & Characterization

The stress-optic modulator consists of a silicon nitride ring resonator and a monolithically integrated PZT actuator, as shown in Fig. 2(a). The ring resonator has a radius of 625 μm and is fabricated using the silicon nitride ultra-low loss CMOS-compatible waveguide process described in [47]. The 175 nm thick LPCVD deposited $Si_3N_4$ waveguide core is sandwiched between a 15 μm thermally grown lower cladding oxide on a silicon substrate and a 6 μm thick TEOS-PECVD deposited top cladding oxide, as illustrated in Fig. 2(b). Chemical-mechanical polishing (CMP) is performed on top of the upper layer oxide to planarize the surface. A 500 nm PZT layer with upper and lower platinum (Pt) electrodes and $TiO_2$ adhesion layer are deposited and patterned laterally offset from the waveguide core by 2 μm with respect to its center. The PZT actuator dimensions and waveguide-offset are designed to achieve a large lateral strain effect across the nitride core while minimizing overlap with the optical mode, and therefore minimize optical losses, as indicated in Fig. 2(b) and described in further detail in section 1 of the supplementary materials. A top-down photograph of the fabricated device is shown in Fig. 2(c). Further details of the fabrication process are described in the section 2 of the supplementary materials.

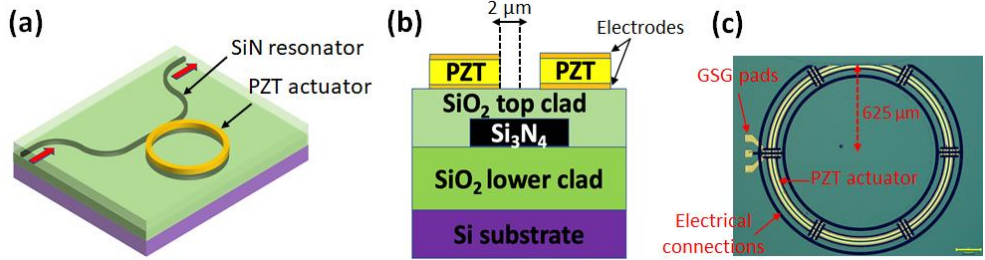

Fig. 2 (a) Illustration of the stress-optic microresonator modulator, the PZT actuator (yellow) is monolithically integrated on top of the Si$_3$N$_4$ resonator without under-etching process. (b) The cross-section of the device, the PZT actuator is placed on top of the waveguide with 2 μm offsets to achieve best performance without affect the optical loss. (c) The optical image of the fabricated device.

The full-width-at-half-maximum (FWHM) of the modulator resonance is measured to be 54.52 MHz at 1550 nm using a radio frequency (RF) calibrated unbalanced MZI with 5.87 MHz free-spectral-range (FSR) as a frequency ruler [48], as shown in Fig. 3(a). The measurement yields a 7.1 million intrinsic Q, a 3.6 million loaded Q, and a corresponding 0.03 dB/cm waveguide loss. Static tuning of the PZT actuator shifts the ring resonance as a function of applied bias voltage is shown in Fig. 3(b). The electric field is applied to the PZT actuator electrodes using a DC probe and strain is induced through the piezoelectric effect which changes the waveguide refractive index. We measure a 14 dB ER across the full 4 GHz tuning range, as the applied voltage is varied from 0 V to 20. A plot of the peak wavelength shift as a function of bias voltage is shown in Fig. 3(c), resulting in a linearly fitted tuning coefficient of -1.6 pm/V or -200 MHz/V. A nonlinearity is observed in the tuning due to hysteresis of the ferroelectric PZT film [49]. The half-wave voltage-length product $V_\pi L$ of the modulator is calculated to be 43 V·cm and $V_\pi L\alpha$ when taking the waveguide loss into account is 1.3 V·dB, which is comparable to the state-of-art phase modulators [39,43,50]. We measure the electrical power consumption to be 20 nW at 20 V bias voltage using a precision source and measure unit (Keysight B2902A) with 100 fA measurement resolution.

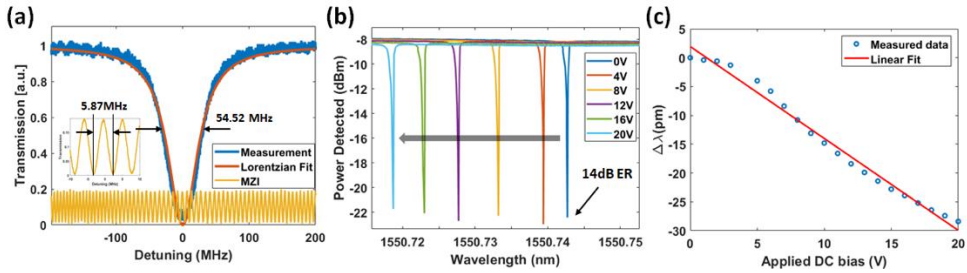

Fig. 3 (a) The Q measurement of the resonator. The calibrated Mach-Zehnder interferometer (MZI) with a 5.87 MHz free-spectral-range (FSR) acts as a frequency ruler (yellow trace) to measure the full-width-at-half-maximum (FWHM) of the resonator resonance (blue trace). (b) The optical transmission spectrum of the static tuning of the device. The resonance has a 14 dB extinction ratio (ER) across the 4 GHz tuning range. (c) The linear fitting of the tuning curve corresponds to -1.6 pm/V or -200 MHz/V tuning coefficient.

The small-signal electrical-to-optical modulation response $S_{21}$ is measured by modulating a semiconductor diode laser (Velocity TLB-6730) tuned to the FWHM point of the resonance , a calibrated fast photodetector (Thorlabs DET01CFC, bandwidth 1.2 GHz) and a vector network analyzer (Keysight N5247B PNA-X). As shown in Fig. 4(a), the 3-dB and 6-dB modulation bandwidths are 20 MHz and 25 MHz respectively. The noise floor is taken by repeating the measurements without PZT modulation. The frequency response bandwidth of the microresonator modulator is primarily limited by the cavity photon lifetime and PZT actuator RC time constant, and can be expressed as [51]:

$$\frac{1}{f_{3dB}^2} = \frac{1}{f_Q^2} + \frac{1}{f_{RC}^2} \quad (1)$$

where $f_Q = \frac{1}{2\pi\tau_{ph}}$ is the frequency limitation by the cavity photon lifetime $\tau_{ph} = Q\lambda/2\pi c$ and $f_{RC} = \frac{1}{2\pi RC}$ is the frequency limitation by the RC time constant of the device. The cavity photon lifetime limited bandwidth corresponds to the resonator linewidth and the main limiting factor to modulation bandwidth is the high resonator Q. The $S_{21}$ response has spikes at higher frequencies that are the mechanical resonances due to vertically propagating acoustic modes. These modes can be reduced by roughening the substrate backside to flatten the frequency response and increase the modulation bandwidth [41].

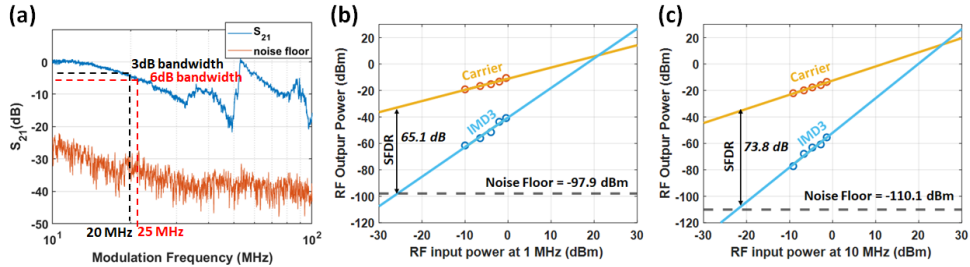

Fig. 4 (a) The frequency response ($S_{21}$) indicates that the 3-dB and 6-dB modulation bandwidth is 20 MHz and 25 MHz, respectively. (b) and (c) RF output power of the fundamental and third order intermodulation distortion (IMD3) component as a function of RF input power of the ring modulator at 1 MHz and 10 MHz.

We characterize the modulator linearity for application such as PDH locking, by measuring the IMD3 SFDR [52–54]. The SFDR is defined as the signal level at which the noise floor and power of the third order distortion tone are equal. The two-tone test is performed by applying two closely placed RF signals to the modulator and measuring the intermodulation components at the optical output using an electrical spectrum analyzer (ESA). The measured noise floor is -97.9 dBm/Hz at 1 MHz and -110.1 dBm/Hz at 10 MHz. IMD3 SFDR values of 65.1 dB·Hz$^{2/3}$ and 73.8 dB·Hz$^{2/3}$ are measured at 1 MHz and 10 MHz, respectively (Fig. 4 (b) and (c)). Further details of the SFDR measurement are described in the section 4 of the supplementary materials.

## 3. Control applications

Two control applications are demonstrated, PDH stabilization of a laser to an optical reference cavity [22] and laser carrier tracking [55]. Figure 5(a) shows the PDH locking experimental setup. The PZT modulator is used in place of an EOM that is typically used in PDH stabilization, to generate double sidebands on the laser carrier. An AOM is used to frequency shift the DSB modulated carrier and lock it to a reference cavity resonance. The reference cavity is an integrated silicon nitride bus-coupled resonator with an ultra-high Q of $10^8$ and large mode volume of $2.9\times10^6$ μm$^3$ [56]. The narrow resonance provide by the high Q is utilized to suppress the laser frequency noise fluctuations and the large mode volume is chosen to reduce the thermo-refractive noise (TRN) floor.

The modulation depth of the DSB modulated carrier is an important consideration for laser noise stabilization. In this experiment, a 20 MHz, 8 V peak-to-peak sinusoidal voltage is applied to the PZT actuator (Fig. 5(b)). Amplitude modulation produces a modulation depth that can be approximated by:

$$A(t) = [1 + m_0 \cos(\Omega t)]A_0 e^{i\omega t} = A_0 e^{i\omega t} + \frac{m_0 A_0}{2} e^{i(\omega+\Omega)t} + \frac{m_0 A_0}{2} e^{i(\omega-\Omega)t} \quad (2)$$

The sideband modulation depth at $\omega \pm \Omega$ (Fig. 5(c)) is calculated to be $m_0 = 0.48$ by fitting Eqn. 2 to the measured response with the sideband-to-carrier power ratio $\frac{m_0^2}{2} = 12\%$. The laser frequency noise (FN) is measured using the self-delayed homodyne laser frequency noise method with an unbalanced fiber MZI optical frequency discriminator (OFD) [57] with an FSR of 1.03 MHz. The power spectral density of the frequency noise as a function of frequency offset from carrier is shown in Fig. 5(d). The stabilized laser frequency noise (red curve) is reduced by four orders of magnitude (~40 dB) compared to the unstabilized laser over the frequency range 100 Hz to 1 kHz and reduced by two orders of magnitude (~20 dB) at 10 kHz frequency offset. The laser stabilization locking circuits has a bandwidth of ~1 MHz where the servo bump can be seen in the stabilized frequency noise measurement. The TRN floor is calculated for the silicon nitride high Q resonator as shown in the green dashed curve in Fig. 5(d). The stabilized laser is able to achieve close to the TRN limit for this cavity over the frequency range of 1 kHz to 10 kHz.

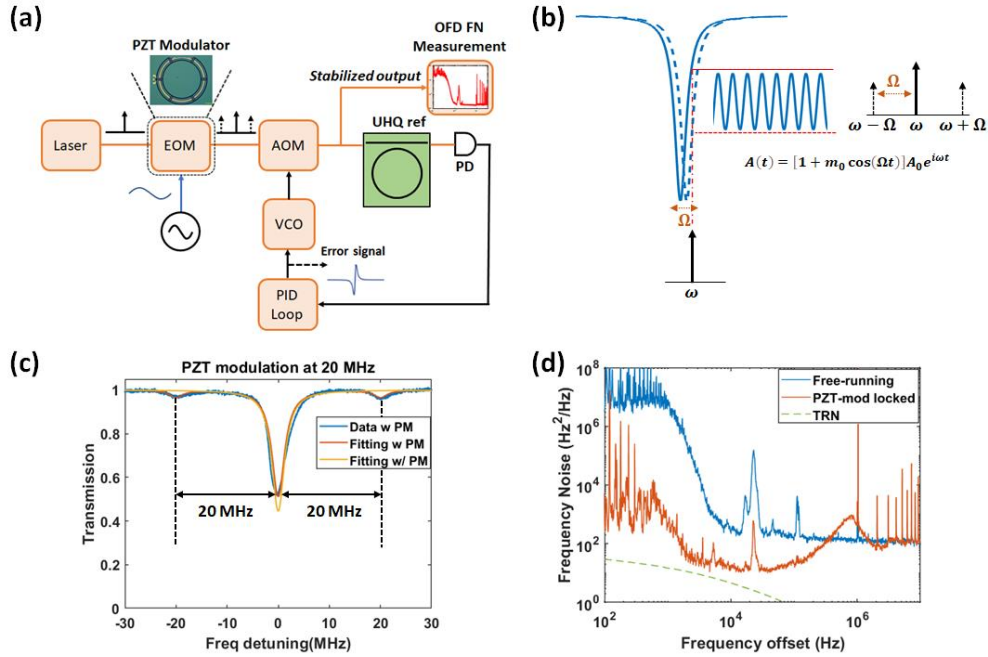

Fig. 5 Laser stabilization demonstration. (a) A semiconductor laser is PDH locked to an ultra-high Q (UHQ) reference cavity using the PZT modulator as a double sideband (DSB) modulator (typically done using an electro-optic modulator (EOM)). The double sideband modulated carrier is frequency shifted by an acousto-optic modulator (AOM) and locked to the quadrature point of an UHQ integrated reference resonator using a proportional–integral–derivative (PID) control loop that drive a voltage-controlled oscillator (VCO) AOM frequency shift control signal. The resulting stabilized frequency noise is measured using an unbalanced fiber MZI optical frequency discriminator (OFD). (b) Illustration of DSB modulated carrier. (c) Measured DSB spectrum of the laser signal with and without the PZT DSB modulation, showing the 40 MHz sidebands with modulation depth of 0.48. (d) Measured frequency noise (FN) power spectral density for the free-running (blue trace) and stabilized laser (red trace) showing 40 dB close to carrier noise reduction and near thermo-refractive noise (TRN) limited performance for the 1 kHz to 10 kHz frequency range. The simulated TRN limit for this UHQ reference cavity is shown in the dashed green trace. PD, photodetector.

Demonstration of the PZT modulator as an automated laser carrier tracking filter is shown in Fig. 6. Laser carrier tracking is important for monitoring and stabilizing wavelength changes to minimize the wavelength drift and spectral misalignments which cause power loss and signal distortion in fiber communications [55]. In this experiment, the stress-optic modulator resonance is locked to the laser output using a PDH locking circuit, as shown in Fig. 6(a). The PDH error signal (shown in the inset) indicates the deviation between the resonator resonance and the laser tone. The PDH servo uses the error signal to control the PZT actuator to lock the resonance to the laser carrier. To demonstrate the tracking function, we measure the filter output in response to a sinusoidal varying output wavelength and to a step response output wavelength shift of the tunable laser. The sinusoidally varying or step input signal ($V_{in}$) is applied to the optical frequency modulation control input of a Velocity TLB-6730 laser. The small signal frequency response of the tracking loop, including the individual responses of the laser, the PZT actuator, the photodetector and the PID loop, is characterized by $V_{out}/V_{in}$ as shown in Fig. 6(b). The bandwidth of the system, characterized at the 180° phase lag point, is $f_{180°} = 0.9$ MHz and is mainly limited by tunable laser bandwidth wavelength control of approximately 1 MHz. As shown in right-side of Fig. 6(b), within the lock bandwidth, the optical level at the tracking filter output of the signal is maintained at a constant value (within 3% at 1 kHz and 10 kHz) with the external signal dithering. When the applied signal reaches 1 MHz, which is near a bandwidth resonance, the optical output of the filter fluctuates. The tracking system step response is measured and shown in Fig. 6(c), where the blue trace is the applied step signal (1 kHz square waveform) to the laser frequency tuning, and the red trace is the control loop response measured at the input to the PZT actuator. The tracking filter output shows the stabilization time (90% to 10%) in reponse to the step wavelength change at the input. The time to stabilize the lock is approximately 130 μs and the settling time is approximately 8 μs.

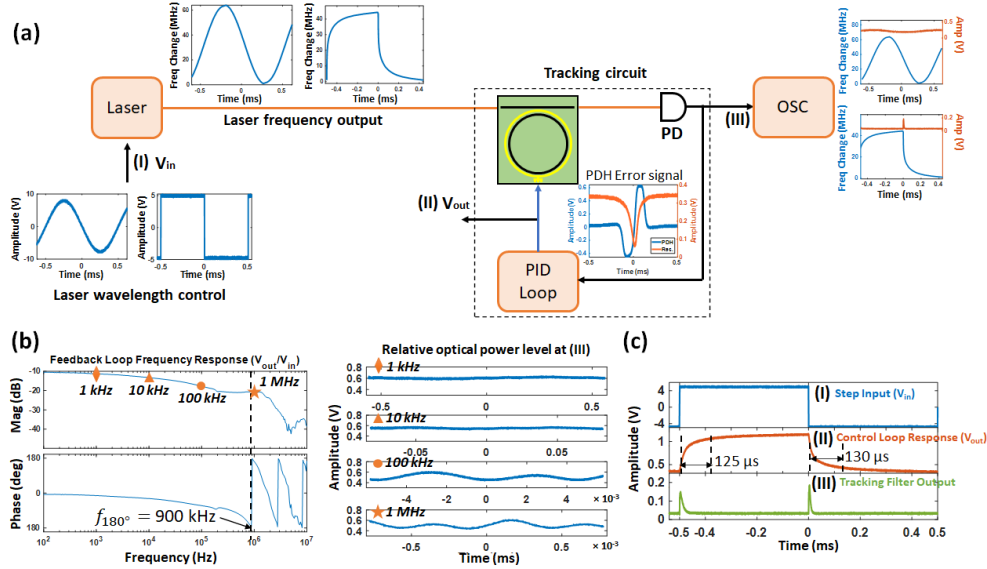

Fig. 6 (a) The PZT-actuated ring modulator is locked to the laser in a PID locking loop, when an external signal is applied to the laser, the locked modulator will respond to the signal dithering and track the laser carrier. The scope at the output port recorded the optical level fluctuation (orange) with the signal dithering in the laser signal (blue). (b) The total response of the feedback loop shows the bandwidth is close to 1 MHz. (c) Step input, control loop response, and tracking filter output stabilization time (from 90% to 10%) ~ 130 μs with ~ 8 μs settling time.

## 4. Discussion

We report a photonic integrated optical modulator for control applications based on a stress-optic PZT actuated ultra-high Q silicon nitride waveguide microresonator. The modulator is a fully planar design, fabricated in an ultra-low loss silicon nitride wafer-scale, CMOS-compatible integration platform [1]. We demonstrate 0.03 dB/cm loss in a 625 μm radius ring operating at 1550 nm, an order of magnitude improvement in modulation bandwidth (20 MHz 3-dB), an order of magnitude decrease in power consumption of 20 nW, a Q of 7.1 million and 14 dB ER across the full tuning range, and tuning efficiency of 200 MHz/V. The PZT design is offset from the waveguide core and optical mode, resulting in a resonator Q and modulation extinction ratio that are relatively constant over the entire 4 GHz tuning range. We also characterize the modulator linearity, an important parameter for many control applications, measuring 65.1 dB·Hz$^{2/3}$ and 73.8 dB·Hz$^{2/3}$ third-order intermodulation distortion spurious free dynamic range at 1 MHz and 10 MHz respectively. These results are presented in Table 1 along with comparable integrated ring modulators based on electro-optic and stress-optic actuation. Our modulator provides an order of magnitude lower power consumption with ultra-high Q, low loss, and ER, V$_\pi$L$_\alpha$ and a tuning sensitivity consistent with control applications. Looking forward, the modulation bandwidth can be further improved by optimizing the resonator loss, the waveguide coupler, and actuator design, in order to increase both the cavity photon lifetime limited and RC time constant limited cutoff frequencies. The tuning strength can be improved using a thicker PZT film as well as a thinner oxide cladding.

Table 1. Summary of this work with comparable integrated ring-resonator modulators

| Modulator | Actuation | BW | Loss | V$_\pi$L$_\alpha$ | Tuning | Q | ER | Power |
|---|---|---|---|---|---|---|---|---|
| SiN + LN / Ring [38] | LN/EO | NA | 0.32 dB/cm | 9.8 V·dB | 1.78 pm/V | Q$_{in}$ = 185 K | 27 dB | NA |
| LN / Ring [50] | LN/EO | 30 GHz | 3 dB/cm | 5.4 V·dB | 7 pm/V | Q$_L$ = 50 K | 10 dB | NA |
| AlN / Ring [58] | AlN/EO | 2.3 GHz | 0.6 dB/cm | 61 V·dB | 0.53 pm/V | Q$_L$ = 800 K | 10 dB | NA |
| SiN + PZT / Ring [39] | PZT/EO | 33 GHz | 1 dB/cm | 3.3 V·dB | 13.4 pm/V | Q$_L$ = 2.2 K | 10 dB | NA |
| SiN / MZI [42] | PZT/SO | 629 kHz | NA | NA | NA | NA | 25 dB | 300 nW |
| SiN / Ring [43] | PZT/SO | <1 MHz | 0.3 dB/cm | 1.1 V·dB | 25.75 pm/V | Q$_L$ = 86 K | 15 dB | 160 nW |
| SiN / Ring [45] | AlN/SO | 91.71 MHz | 0.02 dB/cm | NA | 0.12 pm/V | Q$_{in}$ = 15 M | NA | 300 nW |
| **SiN / Ring (This work)** | **PZT/SO** | **20 MHz** | **0.03 dB/cm** | **1.3 V·dB** | **1.59 pm/V** | Q$_L$ = **3.5 M**$^\dagger$ <br> Q$_{in}$ = **7 M**$^\dagger$ | **14 dB**$^\dagger$ | **20 nW** |

EO: electro-optic modulation; SO: stress-optic modulation. $^\dagger$Across the 4 GHz tuning range

The PZT actuator was used in two application demonstrations, laser PDH stabilization and laser carrier tracking. Laser stabilization improved the close to carrier frequency noise of a laser by 4 orders of magnitude using an ultra-high Q resonator of the same nitride platform. The compatibility of this planar modulator with other silicon nitride devices, e.g., filters, Brillouin lasers, stabilization cavities, will lead to the next level of functions on-chip. Tracking circuits are useful when aligning modulators to wavelength-division multiplexing (WDM) transmitter channels, for filtering a channel or aligning add/drop filters for a WDM transmission system or network. Other applications that will benefit from photonic integration with this modulator include frequency comb control [45], and tunable stimulated Brillouin scattering (SBS) [59]. Since the modulation is based on the stress-optic effect in the silicon nitride waveguide core, and the optical mode does not overlap with the actuators, the modulator design can be transferred to the visible range with a simple waveguide design change (thickness and width), and maintain the low waveguide loss properties for AMO applications such as strontium transitions in the 461 nm to 802 nm wavelength range [3] and rubidium transitions in the 780 nm range as well as other atomic and molecular species with transitions across the nitride transparency window (405 nm – 2350 nm). The wide optical transparency window enabled by the stress-optic modulation extends the application with low-loss silicon nitride integration to the visible [29] for atomic clock and quantum applications. In general, atomic, molecular and optical applications that make use of power consuming bulk-optic modulators, frequency shifters, and other components, will benefit from control modulator integration.

**Funding.** This material is based upon work supported by the National Science Foundation under EAGER Grant No. 1745612, the Advanced Research Projects Agency-Energy (ARPA-E), U.S. Department of Energy, under Award Number DE-AR0001042 and the Army Research Laboratory under Cooperative Agreement Number W911NF-22-2-0056. The views and conclusions contained in this document are those of the authors and should not be interpreted as representing the official policies, either expressed or implied, of the Army Research Laboratory or the U.S. Government.

**Acknowledgments.** The authors would like to thank Steven Isaacson of General Technical Services and Karl Nelsen at Honeywell for their efforts in fabrication of the devices. The authors would like to thank Jane Sabitsana-Nakao at Keysight Technologies for help with the testing instrumentation.

**Disclosures.** The authors declare no conflicts of interest.

**Data availability.** The data that support the findings of this study are available from the corresponding author upon reasonable request.

**Supplemental document.** See Supplement 1 for supporting content.

# Silicon nitride stress-optic microresonator modulator for optical control applications:

## Supplemental document

This document provides additional information on PZT modulator design and simulation, fabrication process and measurements.

### 5. PZT Actuator design and simulation

The PZT actuator (PZT film and platinum electrodes) is placed at a lateral (horizontal) offset with respect to the optical waveguide center and on top of the upper cladding. This placement maximizes the stress-optic effect and minimizes overlap with the optical mode in order to preserve the low optical loss (Fig. S1. (a)). We use COMSOL Multiphysics to simulate the stress distribution, total displacement and effective mode index change due to the stress-optic effect (Fig. S1. (a) and (b)). The effective index change as a function of PZT offset and thickness is summarized in Fig. S1. (c), indicating that 2 μm offset results in the largest index change. Fabricated resonators with different PZT offsets are tested for Q and loss, as shown in Fig. S1. (d). When the PZT actuator is located directly above the waveguide, the Q and the loss are reduced by approximately 20% compared to without the PZT actuator. With the PZT actuator laterally offset by 2 μm from the waveguide and fabricated on top of the upper oxide cladding, the reduction in loss and Q are under a few tenths of a dB.

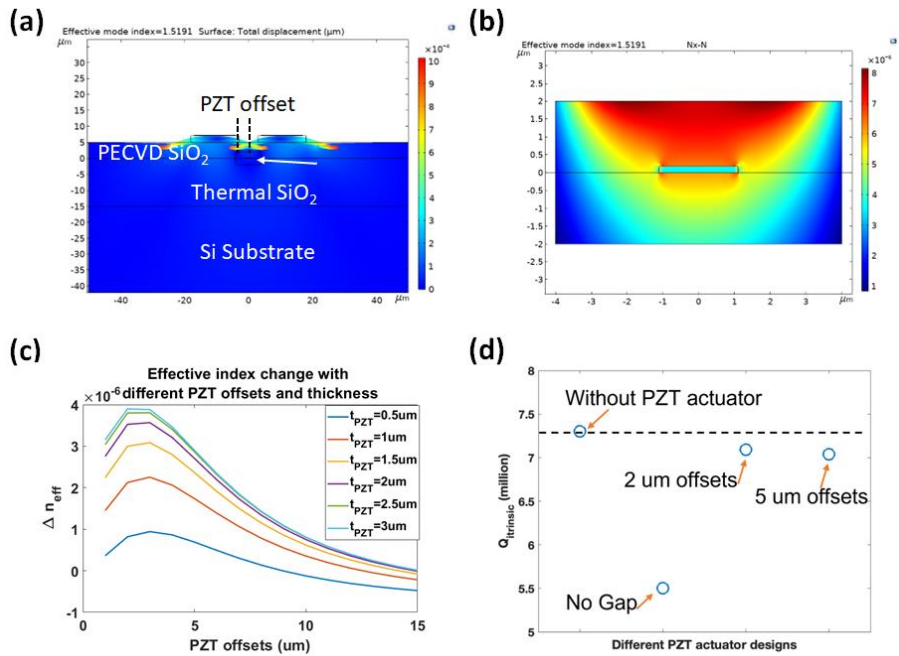

Fig. S1. COMSOL simulation and design of PZT actuator. (a) The total displacement simulation. (b) The effective mode index simulation with different PZT actuator offsets, widths and thickness. (c) Summary of the simulation data, showing that the 2 μm offset results in the largest strain effect. (d) The measurement of the fabricated devices agrees with the simulation.

## 6. Fabrication Process

The waveguides are fabricated on a 15 μm thick lower cladding layer consisting of thermal silica dioxide fabricated on a 1 mm thick, 100 mm diameter silicon substrate. A 175 nm thick stoichiometric silicon nitride layer is deposited by low-pressure chemical vapor deposition (LPCVD) on top of the thermal oxide layer. The waveguides are patterned in the nitride layer using a photoresist mask by 248 nm DUV stepper lithography and inductively coupled plasma etch with CF4/CHF3/O2 gas. Following the pattern etch, 6 μm thick of SiO2 is deposited in two 3-μm steps as the upper cladding layer by plasma enhanced chemical vapor deposition (PECVD) with tetraethylorthosilicate (TEOS) used as a precursor. Next, the wafer is annealed at 1050 °C for 7 hours and 1150 °C for 2 hours. After upper cladding deposition, the wafer is chemical-mechanical polished (CMP) to planarize the surface for the piezo actuator deposition. The actuator stack consists of a sputtered 40 nm thick TiO2 adhesion layer, a sputtered 150 nm thick Pt bottom electrode, and a 500 nm thick layer of PZT (52/48 Zr/Ti ratio) deposited via chemical solution deposition and the stack is capped with a sputtered 100 nm thick Pt top electrode. The PZT and Pt electrodes are patterned by argon ion milling. With the actuator patterned, electrical traces are evaporated and patterned through lift-off and consist of a Cr/Pt/Au stack with thicknesses of 20 nm, 20 nm, and 730 nm respectively. To reduce the resistivity and minimize gold coverage on the electrodes a 10 μm thick copper layer is electroplated using photoresist molds and a sputtered copper seed layer. The photoresist molds and copper seed layer are removed with solvents to release the device.

## 7. Frequency noise measurements

We measure the frequency noise of the laser signal using an optical frequency discriminator (OFD) consisting of a fiber based unbalanced Mach-Zehnder interferometer (MZI) and a balanced photodetector (Thorlabs PDB450C). The frequency noise of the laser, $S_f(v)$, is related to the power spectral density of the detector output $S_{\text{det}}(v)$, expressed as [1]:

$$S_f(v) = S_{\text{det}}(v)\left(\frac{v}{\sin(\pi v \tau_D)V_{pp}}\right)^2 \quad (2)$$

where $v$ is the frequency offset, $\frac{1}{\tau_D}$ is the FSR of the MZI, $V_{pp}$ is the peak-to-peak value of the detector. A high-speed sampling scope (Keysight Infiniivision DSOX6004A) is used to measure the power spectral density $S_{\text{det}}(v)$ with averaging over 16 traces. The $V_{pp}$ is measured with a ramp signal applied to the MZI.

## 8. Modulation bandwidth and linearity measurements

The PZT modulator modulation bandwidth is characterized using the configuration shown in Fig. S2 (a). A tunable continuous wave (CW) laser (Velocity TLB-6730) is tuned to the FWHM point of the resonator resonance and the optical response output is input to a Thorlabs DET01CFC photodetector with bandwidth of 1.2 GHz. The small signal electrical-to-optical response $S_{21}$ is measured with a Keysight N5247B PNA-X vector network analyzer connected to the PZT actuator electrodes and the Throlabs photodetector.

The spurious-free dynamic range (SFDR) measurement setup is shown in Fig. S2 (b). The optical signal from a tunable CW laser (Velocity TLB-6730) is coupled into the PZT modulator which is biased at the FWHM point of the resonator resonance. A two-tone signal input at 1 MHz and 10 MHz is generated using a Keysight EDU33212A waveform generator. The two-tone signal is applied to the modulator with an RF probe and the fundamental and the third-order intermodulation distortion (IMD) frequencies are measured with a photodetector (Thorlabs PDB470C, Bandwidth DC – 400 MHz) and a 40 GHz RF spectrum analyzer (Rohde & Schwarz FSEK20, Bandwidth 9 kHz – 40 GHz). The spectrum of a 10 MHz signal with -4.73 dBm RF input power is shown as an example in Fig. S2 (c) with a measured -80.1 dBm noise floor. The noise power per unit bandwidth is calculated with a 1 kHz resolution bandwidth to be -110.1 dB/Hz assuming the noise floor used in the SFDR calculation [2].

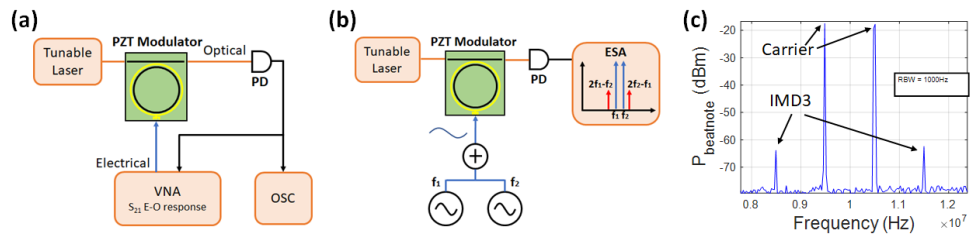

Fig. S2. (a) The experimental diagram for small-signal electrical-to-optical frequency response measurements. (b) The experimental diagram for two-tone spurious-free dynamic range (SFDR) measurements. (c) Detected RF power of the carrier frequencies at 9.5 MHz and 10.5 MHz and the third-order intermodulation distortion (IMD3) frequencies at 8.5 MHz and 11.5 MHz with 1 kHz resolution bandwidth.